\documentclass[prb,aps,superscriptaddress,twocolumn,showpacs]{revtex4-1}
\usepackage{graphicx,amsfonts,bm,amsmath}
\newcommand{\kk}{{\bm k}}
\newcommand{\qq}{{\bm q}}
\newcommand{\rl}{\rangle\!\langle}
\newcommand{\rc}{\rangle^{\mathrm corr}}

\newcommand{\sk}{s_\kk}
\newcommand{\skp}{s_\kk^{(+)}}

\begin{document}

\author{A. Grodecka-Grad}
\altaffiliation[Present address: ]{QUANTOP, Danish National Research Foundation Center for Quantum Optics, Niels Bohr Institute, DK-2100 Copenhagen \O{}, Denmark}
\email{anna.grodecka-grad@nbi.dk}
\affiliation{Computational Nanophotonics Group, Theoretical Physics, 
University of Paderborn, 33098 Paderborn, Germany}
\author{J. F{\"o}rstner}
\affiliation{Computational Nanophotonics Group, Theoretical Physics, 
University of Paderborn, 33098 Paderborn, Germany}

\title{Theory of phonon-mediated relaxation in doped quantum dot molecules}

\begin{abstract}
A quantum dot molecule doped with a single electron in the presence of diagonal and off-diagonal 
carrier-phonon couplings is studied by means of a non-perturbative quantum kinetic theory. 
The interaction with acoustic phonons by deformation potential and piezoelectric coupling is taken into account.
We show that the phonon-mediated relaxation is fast on a picosecond timescale and is dominated
by the usually neglected off-diagonal coupling to the lattice degrees of freedom
leading to phonon-assisted electron tunneling.
We show that in the parameter regime of current electrical and optical experiments, 
the microscopic non-Markovian theory has to be employed.
\end{abstract}


\maketitle

\section{Introduction}

A promising way of building feasible quantum computer is the control of quantum superpositions
in coupled quantum dots (QDs) called quantum dot molecules (QDMs) \cite{fujisawa98}.
In particular, the two electron ground states are used as the logical qubit states
and can be manipulated by electrical \cite{petta05} as well as by optical means \cite{bayer01,hakan07}.
However, the coherent control of the qubit states is limited not only 
by the electron-electron interaction \cite{golubev98}
but also by the coupling to the semiconductor host materials resulting 
in emission and absorption of phonons \cite{brandes99,debald02}.

Since the energy difference between the two electron ground states is of order of a few meV,
this can lead to new phonon-mediated relaxation processes \cite{schinner09}.
The electron transfer between the QDs is of great importance for the currently
performed optical and transport experiments \cite{stinaff06,barthel09}, but only little theoretical work
exists on the influence of the phonon coupling. 
It was mostly limited to perturbative and Markovian theory
or by including only diagonal electron-phonon interaction \cite{brandes99,rozbicki08,climente06,grodecka08}.
A comprehensive theory including non-Markovian processes and non-diagonal coupling 
to the phonon reservoir is required to adequately describe this important solid state qubit system. 

In this paper, we present a fully quantum kinetic description of the phonon-mediated
relaxation in doped quantum dot molecules including non-Markovian effects. 
We show that the phonon-electron interaction in coupled quantum dots
leads to a fast electron relaxation on a picosecond timescale and strongly affects the coherent electron evolution.
A non-perturbative theory based on correlation expansion technique \cite{rossi02,forstner03}
including up to three particle correlations is used, which describes up to two-phonon assisted processes.
We include both diagonal and off-diagonal electron-phonon couplings and show that the usually neglected
off-diagonal contribution can play a dominant role in relaxation processes
leading to fast phonon-assisted electron tunneling.
We derive a compact expression for the electron occupation and coherence evolution in the Markovian approximation
and compare the results with the full quantum kinetic description.

The paper is organized as follows. In Sec.~\ref{sec2}, a quantum kinetic model describing 
an electron confined in a quantum dot molecule coupled to the phonon environment is introduced. 
Next, Sec.~\ref{sec3} contains the results on phonon-assisted electron evolution
and the comparison between non-Markovian and Markovian regime.
In Sec.~\ref{sec4}, we conclude the paper with final remarks. 
In addition, some technical details on the correlation expansion technique are presented in the Appendix.

\section{Model system}\label{sec2}

The system under study consists of a singly doped quantum dot molecule build from two quantum dots.
We consider the relaxation between the two energetically lowest states of the electron
$|1\rangle$ in the left and $|2\rangle$ in the right quantum dot. 
The free electron Hamiltonian reads
\begin{equation}
H_{\mathrm{c}} = \epsilon \left( |2\rl 2| - |1\rl 1| \right) + \Gamma \left(|1\rl 2| + |2 \rl 1 | \right),
\end{equation}
where $\Gamma$ is the tunneling coupling between the QDs and $2\epsilon$ is the energy difference
between the ground states in both quantum dots.
The free phonon Hamiltonian reads $ H_{\mathrm{ph}} = \sum_{\kk,s} \hbar \omega_{\kk,s} b_{\kk,s}^\dag  b_{\kk,s}$
with phonon creation and annihilation operators $b_{\kk,s}^\dag$ and $b_{\kk,s}$, respectively,
phonon frequencies $\omega_{\kk,s}$ and phonon wave vector $\kk$, where $s$ labels different phonon branches
(longitudinal $l$ and two transversal $t1$ and $t2$).
The carrier interacts with the phonon reservoir, which is described by the following Hamiltonian
\begin{equation}
H_{\mathrm{int}} = \sum_{i,j=1,2} \sum_{\kk,s} \left[ g_{ij,s} (\kk) |i\rl j| b_{\kk,s} + 
g_{ij,s}^* (\kk) |j\rl i| b_{\kk,s}^\dag \right],
\end{equation}
where $g_{ij,s} (\kk)$ are coupling elements of the electron-phonon interaction. 
Note, that we take into account the usually neglected off-diagonal electron-phonon coupling,
describing the direct phonon-mediated electron transfer between the quantum dots with 
simultaneous emission or absorption of a phonon, which, as we will show, leads to a fast relaxation channel.

We consider the relevant acoustic phonons coupled via both piezoelectric coupling and
deformation potential with corresponding coupling elements:
\begin{equation}
g_{ij,s}^{\mathrm{PE}}(\kk) = -i \sqrt{ \frac{\hbar}{2 \rho c_s k} } 
\frac{d_P e}{\varepsilon_0 \varepsilon_r} M_s(\hat k)
\int d^3r \; \psi_i^*({\bf r}) e^{i\kk{\bf r}} \psi_j ({\bf r}),
\end{equation}
\begin{equation}
g_{ij,l}^{\mathrm{DP}}(\kk) = \sqrt{ \frac{\hbar k}{2 \rho V c_l} }  D_{e} 
\int d^3r \; \psi_i^*({\bf r}) e^{i\kk{\bf r}} \psi_j ({\bf r})
\end{equation}
with Gaussian electron wave functions
\begin{equation}
\psi_i({\bf r}) = \frac{1}{\pi^{3/4} l_i^{3/2}} \exp\left({-\frac{{\bf r}^2}{2 l_i^2}}\right).
\end{equation}
In the numerical simulations, the parameters corresponding to self-assembled GaAs quantum dots have been used:
$l_1 = 4$~nm, $l_2 = 4.1$~nm, and $d = 6$~nm, where $l_i$ is the electron wave function size of the $i$-th QD, and the distance between the dots is~$d$.
Here, $\rho = 5360$~kg/m$^3$ is the crystal density, $V$ is the normalization volume of the phonon modes,
$c_s$ is the speed of sound (longitudinal $c_l = 5150$~m/s or transverse $c_t = 2800$~m/s, 
depending on the phonon branch),
$d_P = 0.16$~C/m$^2$ is the piezoelectric constant, $D_e = -8$~eV is the deformation potential constant 
for the electrons and $\varepsilon_r = 13.2$ is the static dielectric constant. 
The function $M_s(\hat k)$ in the piezoelectric coupling element depends 
only on the orientation of the phonon wave vector
and for a zinc-blende structure reads: $M_l(\hat k) = \frac{3}{2} \sin(\theta) \sin(2\theta) \sin(2\varphi)$,
$M_{t1}(\hat k) = - \sin(2\theta) \cos(2\varphi)$, and 
$M_{t1}(\hat k) = \sin(\theta) [3\cos^2(\varphi)-1] \sin(2\varphi)$.

In order to study the dynamics of the electron confined in a quantum dot molecule
in the presence of electron-phonon interaction, 
a second order correlation expansion method is used \cite{rossi02,forstner03}.
This is a non-perturbative technique which can be employed for the microscopic quantum kinetic
description of the dynamics covering the non-Markovian memory effects
(see Appendix for details).
The equations of motion for the electron occupations $\langle |1\rl 1| \rangle$, $\langle |2\rl 2| \rangle$,
coherences $\langle |1\rl 2| \rangle$, and the phonon-assisted elements
e.g. $\langle |1\rl 2| b_{\qq,s} \rc$ are derived and solved by standard numerical methods.
We included up to three particle correlations, e.g. $\langle |1\rl 2| b_{\qq,s}^\dag b_{\kk,s'} \rc$,
which leads to a description of up to two-phonon assisted processes.
An alternative approach to study the phonon-assisted dynamics is the
real time renormalization-group method \cite{keil02}.

\begin{figure}[b] 
\begin{center} 
\unitlength 1mm
{\resizebox{90mm}{!}{\includegraphics{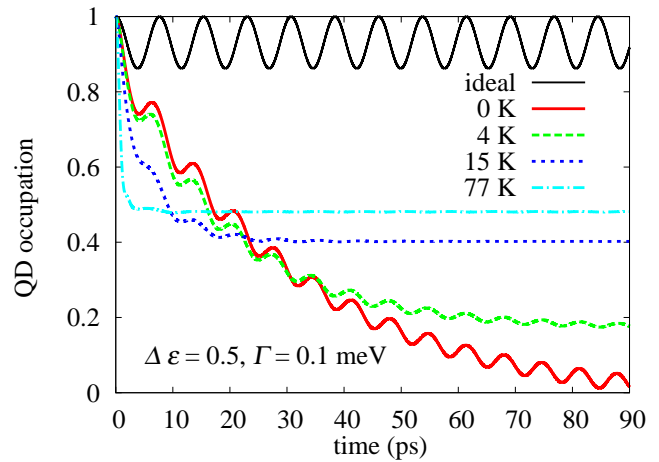}}}
\end{center} 
\caption{\label{fig:1-temp} The occupation probability of the quantum dot with the higher energy 
as a function of time at different temperatures with $\Delta\epsilon = 0.5$~meV, and $\Gamma = 0.1$~meV,
and the ideal electron evolution without coupling to phonons.}
\end{figure}

\section{Phonon-assisted electron evolution}\label{sec3}

The electron is initially in a quantum dot with higher energy, $f = \langle |2\rl 2| \rangle =1$. 
We assume that the injection time is much shorter than the phonon reservoir response,
so that all the correlation terms can be initially set to zero. 
In Fig.~\ref{fig:1-temp}, the occupation of the QD with higher energy with and without coupling
to the phonon reservoir at four different temperatures is shown as a function of time. 
As expected, the ideal electron evolution without interaction with phonons (solid black line) 
is a coherent oscillation of the electron between the two quantum dots with period determined 
by the tunneling coupling and the energy difference between the two electron states $\Delta\epsilon$
and is proportional to $\Gamma/(\Delta\epsilon)$. 
For the presented set of parameters with $\Delta \epsilon = 0.5$~meV and $\Gamma = 0.1$~meV, 
the electron oscillates slightly but mostly stays in the QD with higher energy.
The tunneling coupling is to small to let the electron fully tunnel between the dots.

\begin{figure}[b] 
\begin{center} 
\unitlength 1mm
{\resizebox{90mm}{!}{\includegraphics{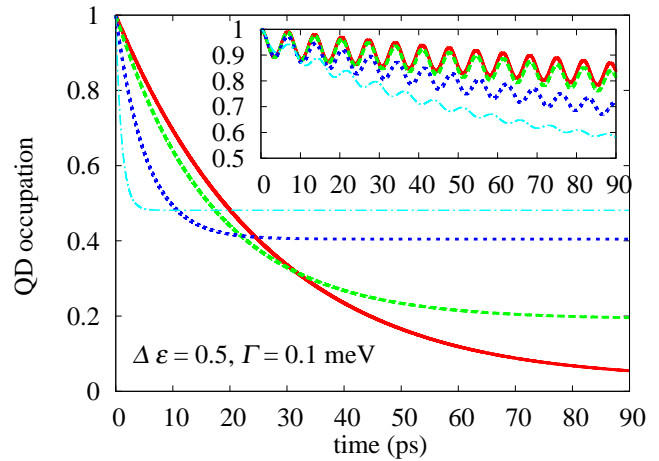}}}
\end{center} 
\caption{\label{fig:m-temp} As in Fig.~\ref{fig:1-temp} but in the Markovian approximation.
Inset: The evolution of the electron occupation including only the diagonal electron-phonon coupling.
The lines and axes as in Fig.~\ref{fig:1-temp}.}
\end{figure}

If the coupling to the phonon environment is included, the dynamics of the confined electron changes completely:
the electron tunnels to the neighboring dot already on the picosecond time scale.
At low temperature, the electron tunnels from the quantum dot with the higher energy 
to the one with the lower energy. At $T=0$~K, only the emission processes are available,
thus, the electron tunnels completely to the lower QD.
As the temperature grows, the probability of the phonon absorption increases. In consequence,
the final state of the electron moves from the quantum dot with the lower energy towards
an equal superposition of electron confined in the left $|1\rangle$ and right $|2\rangle$ quantum dot.
This is due to the fact, that at high temperatures, the probability of phonon absorption
equals to the probability of its emission. 
In addition, the amplitude of the oscillations decreases with growing temperature,
since the damping gets stronger. Similarly, the initial decay is also faster at higher temperature.
We compared the results including up to two-phonon assisted processes to those with up to one phonon
and found out that the former leads only to a small correction.
Thus, the quantum kinetic description with the first order correlation expansion technique
results in the proper description of the electron dynamics.

In order to show that the fast relaxation processes result from the direct phonon-assisted electron tunneling
described by the off-diagonal electron coupling to phonons,
the occupation of a QD modeled using only the diagonal interaction Hamiltonian is shown in the inset 
in Fig.~\ref{fig:m-temp}.
One can see, that in this case the relaxation is considerably slower up to two orders of magnitude.
Thus, in order to properly describe the dynamics of an electron in a quantum dot molecule,
one cannot neglect the off-diagonal electron-phonon coupling. 
A similar conclusion has been met
by Zibik et al. \cite{zibik08,zibik09}, who studied the coherence of the $s$-$p$ transition 
in single quantum dots. It has been shown that both direct absorption of acoustical phonons between
the two $p$ states as well as the scattering via virtual transitions have to be included
in order to explain the decoherence mechanisms of the considered transition. 

Next, for comparison with our full quantum kinetic description, we have derived a Markovian approximation
for the electron dynamics governed by real electron transitions, where the memory effects are neglected.
Including phonon-assisted correlations up to the first order, it is possible to derive compact expressions 
for the electron density and coherence [see Eqs.~(\ref{f}) and~(\ref{p}), respectively].

The first term in the equation of motion for the density $f$ describes the ideal coherent evolution
of the electron, whereas the second term corresponds to the correction due to the electron-phonon coupling.
Here, one can easily recognize the two terms corresponding to the phonon emission $\sim f(n_{\qq,s}+1)$ 
and absorption $(1-f) n_{\qq,s}$ processes, where $n_{\qq,s}$ is the Bose distribution function.
The occupation is directly affected only by the off-diagonal coupling to phonon reservoir.
In contrast to this, the polarization contains also the diagonal elements which lead to pure dephasing processes
\cite{grodecka07,machnikowski04b,grodecka09}.
They are particularly visible in the case of quantum dots with large size difference.
It is worth mentioning that the considered phonon-induced decay is a few orders of magnitude faster 
than the radiative decay of an exciton confined in a single quantum dot, 
which for self-assembled GaAs QDs is in the order of 1~ns \cite{langbein04}.
The results of the numerical simulations for the Markovian approximation are shown in Fig.~\ref{fig:m-temp}.
Here, the coherent oscillations of the electron vanish due to strong diagonal and off-diagonal damping elements.
Thus, the difference between the quantum kinetic description and Markovian approximation
are particularly visible in the case of lower temperatures, where the amplitude of the oscillations
is large in the full description.

\begin{figure}[t] 
\begin{center} 
\unitlength 1mm
{\resizebox{90mm}{!}{\includegraphics{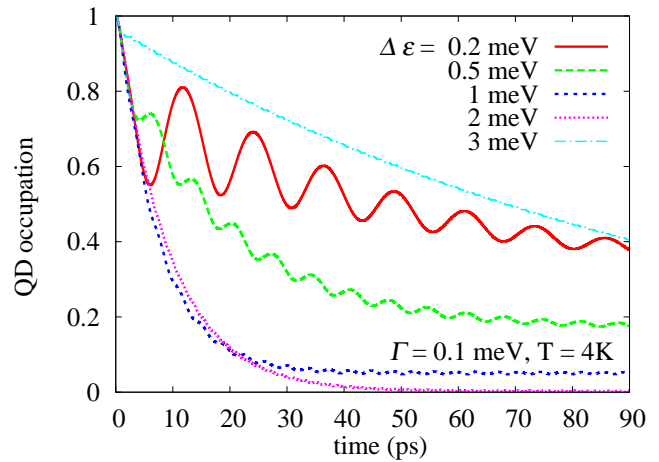}}}
\end{center} 
\caption{\label{fig:1-eps} The time evolution of the electron density in the QD with higher energy
for different energy differences between the electron states with $\Gamma = 0.1$~meV and at $T = 4$~K.}
\end{figure}

As mentioned previously, the tunneling coupling and energy difference determine the period and
amplitude of the coherent oscillations between the two electron states in both quantum dots.
In Fig.~\ref{fig:1-eps}, the time evolution of the QD occupation including the coupling
to phonons is shown for different values of the energy difference $\Delta\epsilon$. 
The energy difference strongly influences the phonon-mediated relaxation,
since the probability of the phonon-assisted processes is high for $\Delta\epsilon \sim $ a few of meV
\cite{grodecka08}. Thus the electron tunneling is very fast for $\Delta\epsilon$ from $0.5$ to $2$~meV.
If the energy difference is larger than the energy cutoff of the phonon density,
which is determined by the size of the quantum dot $\sim c_{\mathrm{s}}/l$, 
the probability of phonon-assisted processes vanishes. 
Thus, for the $\Delta\epsilon=3$~meV the relaxation becomes slower again.
As shown before, the amplitude of the coherent oscillations decreases with growing energy difference
at fixed tunneling coupling $\Gamma = 0.1$~meV and temperature $T=4$~K.

\begin{widetext}
\begin{eqnarray}\label{f}
\dot f & = & \frac{2}{\hbar} \Gamma \mathrm{Im}(p) + \frac{\pi}{\hbar} \sum_{\qq,s} |g_{12,s}(\qq)|^2
 \delta\left(\hbar \omega_{\qq,s} - 2\epsilon \right)  \left\{ 2\left[ f(n_{\qq,s}+1) - (1-f) n_{\qq,s} \right] - 4 \mathrm{Im}^2(p) \right\},\\ \label{p} 
\dot p & = & -\frac{i}{\hbar} 2 \epsilon p -\frac{i}{\hbar} \Gamma (1-2f)
+ \frac{\pi}{\hbar} \sum_{\qq,s} [g_{22,s}^*(\qq) - g_{11,s}^*(\qq)] g_{12,s}(\qq)
 \delta\left(\hbar \omega_{\qq,s} - 2\epsilon \right) 
\left[ f(n_{\qq,s}+1) - (1-f) n_{\qq,s} -|p|^2 \right]  \\ \nonumber
&& + \frac{\pi}{\hbar} \sum_{\qq,s}  [g_{22,s}(\qq) - g_{11,s}(\qq)] g^*_{12,s}(\qq) p^2
 \delta\left(\hbar \omega_{\qq,s} - 2\epsilon \right) 
- \frac{\pi}{\hbar} \sum_{\qq,s} |g_{12,s}(\qq)|^2 (f+n_{\qq,s}) 4i \mathrm{Im}(p)
 \delta\left(\hbar \omega_{\qq,s} - 2\epsilon \right).
\end{eqnarray}
\end{widetext}

\begin{figure}[t] 
\begin{center} 
\unitlength 1mm
{\resizebox{90mm}{!}{\includegraphics{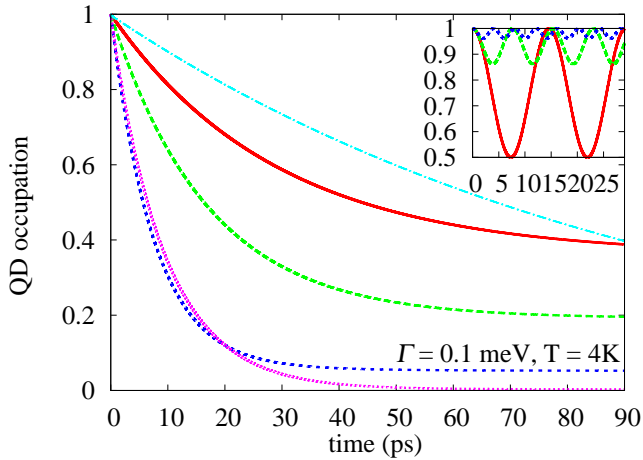}}}
\end{center} 
\caption{\label{fig:m-eps} As in Fig.~\ref{fig:1-eps} but with the Markovian approximation.
Inset: The ideal evolution without interaction with the phonon reservoir.}
\end{figure}

If the Markovian approximation is employed [see Fig.~\ref{fig:m-eps}]
the coherent oscillations vanish, thus the discrepancies from the quantum kinetic theory
is largest for the energy differences approaching the value of the tunneling coupling.
(the ideal electron evolution is shown in the inset in Fig.~\ref{fig:m-eps} for reference).
This case corresponds to the regime which is common in the recently performed 
transport and optical experiments.
Thus, in order to properly describe the electron dynamics, the full non-Markovian description
including the off-diagonal coupling to the lattice degrees of freedom has to be used.
In the case of energy differences much larger compared to the tunneling coupling,
the Markovian approximation delivers the same results as the quantum kinetic description.

\section{Conclusion}\label{sec4}

In conclusion, we have presented a full description of the electron dynamics
in the presence of the electron-phonon coupling in a quantum dot molecule doped with a single electron.
This solid state system is of great importance for the nanoscience, 
in particular for quantum information processing as well as quantum transport and optics.
The two couplings to the relevant acoustic phonons via deformation potential and piezoelectric coupling
have been taken into account.
We showed that the phonon-mediated relaxation is a fast process on a picosecond timescale
and strongly modifies the coherent evolution of the electron. Based on the full quantum kinetic description,
we derived a compact expressions for the electron density and polarization in the Markovian approximation 
and we showed that in the regime of the current electrical and optical experiments, 
the full non-Markovian description is needed.
We showed that the usually neglected off-diagonal electron-phonon coupling is dominant in the electron dynamics
and leads to two orders of magnitude faster relaxation. 

\begin{acknowledgments}
A.\ G.-G. thanks P. Machnikowski and C. Emary for helpful discussions.
A.\ G.-G. and J.\ F.\ acknowledge support from the Emmy Noether Program of the DFG (Grant No. FO 637/1-1)
and the DFG Research Training Group GRK 1464, and
thank the John von Neumann Institut f{\"u}r Computing (NIC) for computing time.
\end{acknowledgments}

\appendix*
\section{Correlation expansion technique}

In this Appendix, we present some details concerning the correlation expansion technique \cite{rossi02,forstner03}, which is based on the assumption that correlations involving
an increasing number of particles are of decreasing importance. 
It is a non-perturbative method widely used for microscopic 
quantum kinetic description of the dynamics including the memory effects in the non-Markovian regime.

Using Heisenberg equation one derives the equations of motion for the quantities of interest:
electron occupation $f = \langle | 2 \rl 2 | \rangle$ and coherence $p = \langle | 1 \rl 2 | \rangle$.
The equation for the time evolution of the electron density
\begin{eqnarray} \nonumber
\dot f & = & \frac{2}{\hbar} T \; \mathrm{Im} (p)
- \frac{i}{\hbar} \sum_\kk g_{12}(\kk) \left[\sk - \sk^{(+)*} + 2i B_\kk \mathrm{Im} (p) \right] \\ \nonumber
&& - \frac{i}{\hbar} \sum_\kk g_{12}^*(\kk) \left[\skp - \sk^{*} + 2i B_\kk^* \mathrm{Im} (p) \right].
\end{eqnarray}
couples to the coherence $p$, $B_{\kk} = \langle b_{\kk} \rangle$, and to the phonon-assisted correlations
$\sk = \langle |1 \rl 2| b_\kk \rc$ and $\skp = \langle |1 \rl 2| b_\kk^\dag \rc$.
Here, the factorization scheme has been used:
$ \langle |1 \rl 2| b_\kk \rangle =  \langle |1\rl 2|\rangle \langle b_{\kk} \rangle +\langle |1 \rl 2| b_\kk \rc$,
where the quantities have been decompose into all possible lower-order factorizations.
The next step is to derive the equations of motion for $p$ and all the phonon-assisted correlation quantities.
The latter couple to three particle correlations, e.g.  $\langle |1\rl 2| b_{\qq,s}^\dag b_{\kk,s'} \rc$
which couple to four particle correlations, etc. Thus, in order to get a closed set of equations,
one needs to truncate the hierarchy by neglecting higher order correlations.
In the present paper, we included up to three particle correlations.
A comprehensive review on the correlation expansion technique is presented in Ref.~\onlinecite{rossi02}.

\end{document}